\documentclass[onecolumn,conference,a4paper]{IEEEtran}

\usepackage{amssymb,amsmath,epsfig,latexsym,graphicx,bm}
\usepackage{rotating}
\usepackage{bm}
\usepackage[colorlinks=true,linkcolor=blue,citecolor=blue,urlcolor=blue]{hyperref}
\usepackage{color}
\usepackage[left = 3cm, top=3cm,right=3cm,bottom=3cm,nohead,nofoot]{geometry}

\usepackage{epstopdf}

\title{Dy doped BiFeO$_3$ : A Bulk Ceramic with
	Improved Multiferroic Properties Compared to Nano Counterparts}

\author{\IEEEauthorblockN{Sayeed Shafayet Chowdhury,$^{1,2}$
		Abu Hena Mostafa Kamal,$^{2}$ Rana Hossain,$^{2}$ \\Mehedi Hasan,$^3$ Md.~Fakhrul Islam,$^3$ Bashir Ahmmad,$^4$ \\and M. A. Basith$^{2,}$\IEEEauthorrefmark{1}}
	\IEEEauthorblockA{$^1$Department of Electrical and Electronic Engineering, \\ Bangladesh University of Engineering and Technology, Dhaka-1000, Bangladesh.\\}
	\IEEEauthorblockA{$^2$Department of Physics, Bangladesh University of Engineering and Technology, Dhaka-1000, Bangladesh.\\}
	\IEEEauthorblockA{$^3$Department of Glass and Ceramics Engineering, \\Bangladesh University of Engineering and Technology, Dhaka-1000, Bangladesh.\\}
	\IEEEauthorblockA{$^4$Graduate School of Science and Engineering, Yamagata University, 4-3-16 Jonan, Yonezawa 992-8510, Japan.\\}
	\IEEEauthorrefmark{1}mabasith@phy.buet.ac.bd}

\begin{document}

	\maketitle

	\begin{abstract}
		The synthesis as well as structural, multiferroic and optical characterization of Dy doped BiFeO$_3$ multiferroic ceramic are presented. Bulk polycrystalline Bi$_{0.9}$Dy$_{0.1}$FeO$_3$ sample is synthesized by solid state reaction, while their nano counterparts are prepared using ultrasonic probe sonication technique. Significant improvement of phase purity in the as synthesized samples is observed after the doping of Dy both in bulk Bi$_{0.9}$Dy$_{0.1}$FeO$_3$ sample and corresponding nanoparticles as evidenced from Rietveld refinement. Magnetization measurements using SQUID magnetometer exhibit enhanced magnetic properties for Dy doped bulk Bi$_{0.9}$Dy$_{0.1}$FeO$_3$ ceramic compared to their nanostructured counterparts as well as undoped BiFeO$_3$. Within the applied field range, saturation polarization is observed for Bi$_{0.9}$Dy$_{0.1}$FeO$_3$
		bulk ceramic only. As a result, intrinsic ferroelectric behavior is obtained just for this sample.
		Optical bandgap measurements reveal lower bandgap for Dy doped bulk Bi$_{0.9}$Dy$_{0.1}$FeO$_3$ ceramic compared to that of corresponding nanoparticles and undoped BiFeO$_3$. The outcome of this investigation demonstrates the potential of Dy as a doping element in BiFeO$_3$ that provides a bulk ceramic material with improved multiferroic and optical properties compared to those of corresponding nanoparticles which involve rigorous synthesis procedure.

	\end{abstract}
\vspace{6mm}

\section{Introduction}
Multiferroic materials have been a core attraction in the research fields of advanced functional and nanostructured materials
in recent times \cite{ref1}--\cite{ref3} since they possess simultaneous ferromagnetic, ferroelectric, and/or ferroelastic orderings. The multiferroic BiFeO$_3$ (BFO) has been a paradigm among them due to its potential in the novel multifunctional devices \cite{ref101,ref4}. The BiFeO$_{3}$ is ferroelectric below T$_{C}$$\sim$1103 K and antiferromagnetic (AFM) below T$_{N}$$\sim$643 K. It is of rhombohedrally distorted perovskite ABO$_{3}$ (A = Bi, B = Fe) structure with space group \textit{R3c} and lattice parameters a = 5.58 $\AA$ and c = 13.87 $\AA$ \cite{ref4b}--\cite{ref4d}. However, the suitability of undoped BFO in many cases becomes constrained as a result of its spiral modulated spin structure (SMSS) \cite{ref5}. Moreover, the formation of different impurity phases \cite{ref6} further deteriorates the properties of undoped BFO. Thus, to overcome these hindrances and to achieve superior multiferroic properties, several attempts have been made by substituting Bi and Fe with rare-earth and transition metal elements, respectively in BFO.  In fact, the potential applications of multiferroic BFO in the field of information storage technology by maneuvering the inherent magnetic and ferroelectric properties using dopants has opened up a new horizon of research. Enhanced multiferroicity of the parent material was achieved by substituting Bi by ions such as Gd, La, Nd etc.~or by simultaneous substitution of Bi and Fe in BFO by ions such as La and Mn, La and Ti, Nd and Sc, Gd and Ti etc., respectively, only a few to mention from numerous examples \cite{ref7}--\cite{ref10}. Most importantly, a significant improvement in magnetic and ferroelectric properties were observed in the undoped and cation doped BFO nanoparticles with particle size less than periodicity of helical order $\sim$62 nm \cite{ref107}---\cite{ref102} compared to that of parent BFO. In previous investigations, we also observed improved structural and magnetic properties for Gd doped and Gd-Ti co-doped BFO nanoparticles compared to their corresponding bulk ceramics \cite{ref11, ref19}.  Withstanding the fact that improved multiferroic properties of undoped and cation doped BFO nanoparticles were reported in various investigations \cite{ref102}---\cite{ref106}, the associated jeopardy was that rigorous synthesis procedure had to be adopted to obtain nanoparticles with sizes less than ~62 nm. Such exhaustive synthesis routes could be avoided if equivalent or more preferably, superior properties could be obtained using the bulk materials, which involve considerably lower synthesis complexity as well as lower cost. Therefore, the search for a bulk ceramic material with comparable or improved multiferroic performance than that of nano counterparts holds significance. In the present investigation, we have synthesized Dy doped bulk Bi$_{0.9}$Dy$_{0.1}$FeO$_3$ ceramic and their corresponding nanoparticles. The structural, multiferroic and optical properties were investigated and compared between bulk ceramic and nanoparticles. 

The rationale behind performing doping of Dy$^{3+}$ ions is threefold: (i) to suppress the impurity phases of BFO, (ii) to modify spiral spin structure of BFO for enhancing magnetization, and (iii) to reduce leakage current density to obtain improved  ferroelectric properties. Due to the size difference between Bi$^{3+}$ (1.03$\AA$) and Dy$^{3+}$ (0.912$\AA$) ions, we anticipate a change in structural properties of bulk ceramic Bi$_{0.9}$Dy$_{0.1}$FeO$_3$ compared to undoped BFO. The
introduction of Dy$^{3+}$ in the Bi-site of BFO is expected to modify its magnetic
properties because of the large magnetic moment of Dy$^{3+}$ ($\sim$ 10.6 $\mu$B) ions
which could result in additional magnetic interactions and ordering \cite{ref10a}.  Moreover, the bond enthalpy of Dy-O (607 $\pm$ 17 kJ/mol) is stronger than that of Bi-O (337 $\pm$ 12.6 kJ/mol) \cite{ref10b}. As a result, reduction in the amount of oxygen vacancies is expected. The doping concentration of Dy has been kept nominally 10\% for which superior multiferroic properties were reported in comparison with other doping concentrations \cite{ref26}--\cite{ref13}. Therefore, 10\% Dy doped BFO bulk sample and their corresponding nanoparticles were synthesized to compare the multiferroic and optical properties between bulk ceramic and nanoparticles. For comparison, BFO bulk ceramic material was also prepared. To
further compare our findings with nanoparticles of identical composition prepared by a different synthesis technique,
some results reported in \cite{ref17} were considered where Dy doped
BiFeO$_3$ nanoparticles were synthesized by a low temperature co-precipitation method. Our investigation demonstrated enhanced structural purity and multiferroic properties in Dy substituted Bi$_{0.9}$Dy$_{0.1}$FeO$_3$ bulk ceramic material compared to those of its corresponding nanoparticles and parent BiFeO$_3$.

\section{EXPERIMENTAL DETAILS}
The polycrystalline samples of undoped  BiFeO$_3$ (referred to as BFO) and Dy doped Bi$_{0.9}$Dy$_{0.1}$FeO$_3$ (referred to as BDFO) were synthesized using standard solid state reaction technique \cite{ref10}. To prepare the bulk polycrystalline samples, the calcined powders were pressed into pellets using a uniaxial hydraulic press and sintered at 825$^o$C for 5 hours at heating rate of 3 $^o$C per minute. The pellets were ground again into powder materials.  The
nanoparticles were fabricated directly from these bulk powder materials using the
ultrasonic probe dispersion technique with varying amount of
sonication time by the application of ultrasonic energy generated from a probe \cite{ref11}. For the synthesis of nanoparticles, a portion of the micro-meter sized powder was subsequently mixed with isopropanol. We then put the mixtures of isopropanol and powder with a mass percentage of $\sim$0.5\%
\cite{ref11} into an ultrasonic probe and
sonicated the solutions for 30, 60 and 90 minutes respectively, for three different samples. After around six hours, $\sim$40\% of
the mass was collected as supernatant and was used
for characterization. X-ray diffraction (XRD) analysis was performed using a diffractometer with CuK$_{\alpha}$  (${\lambda}$ = 1.5418 $\AA$) radiation to study the crystalline structures of the as synthesized materials. The magnetization vs. applied magnetic field (M-H) hysteresis loops of BFO, BDFO bulk and nanoparticles were investigated
using a Superconducting Quantum Interference Device
(SQUID) Magnetometer (Quantum Design MPMS-XL7,
USA). The temperature dependent magnetization measurements were investigated both at zero field cooling
(ZFC) and field cooling (FC) processes. To measure electrical properties of nanoparticles, pellets were prepared by pressing the powders using a hydraulic press and annealing at 750$^o$C with high heating rate (20$^o$C/min) \cite{ref106}. A ferroelectric loop tracer in conjunction with an external amplifier (10 kV) was used to trace the leakage current density and ferroelectric polarization of the pellet shaped samples. Finally, ultraviolet-visible diffuse reflectance
spectra (UV-Vis DRS mode) of the samples were measured by a UV-visible spectrometer (UV-2600, SHIMADZU).

\section{RESULTS AND DISCUSSION}

\subsection{Structural Characterization}
\begin{figure}[hh]	
	\centering
	\includegraphics[width=3.5in]{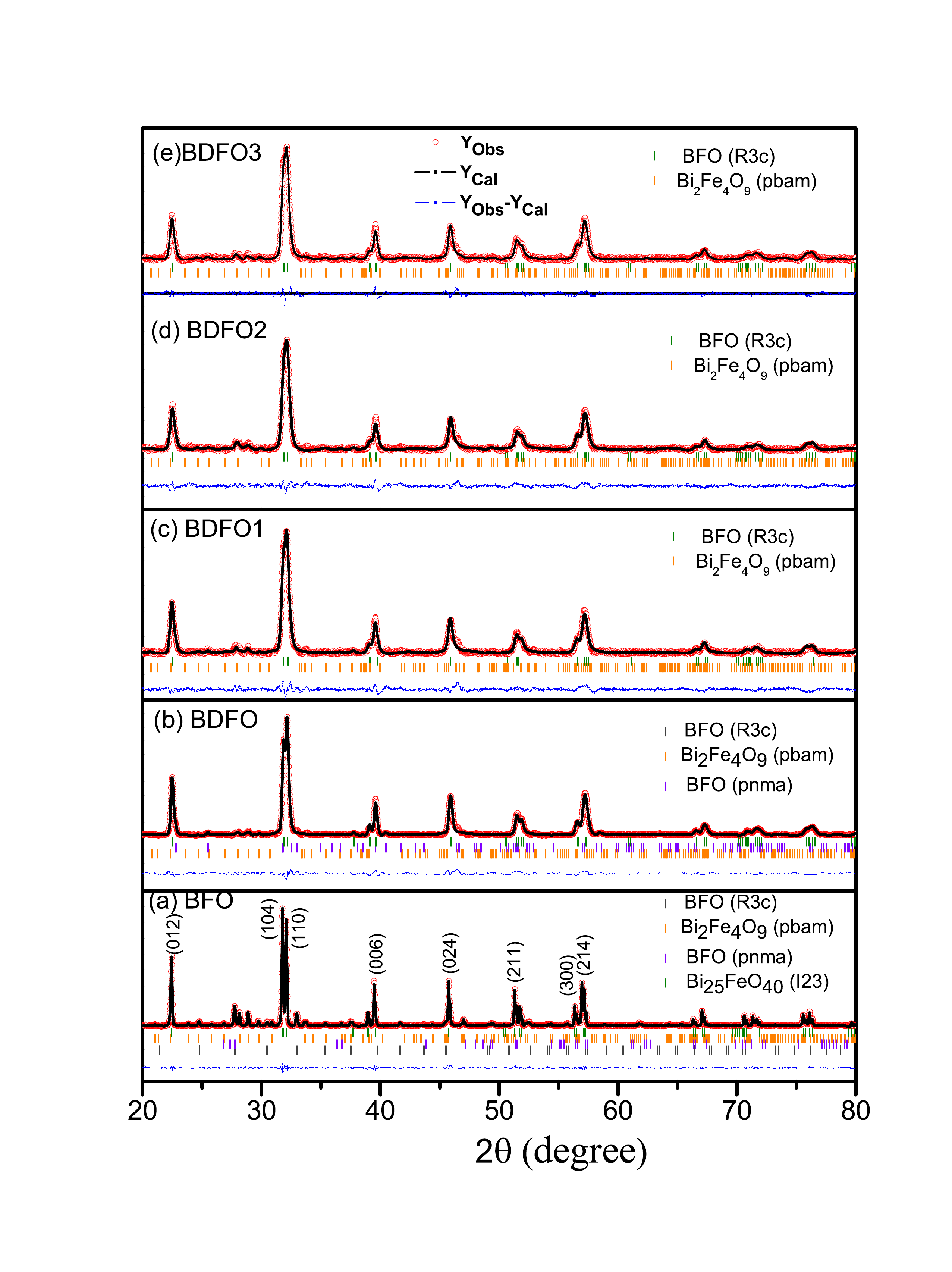}\\
	\caption{Rietveld refined XRD patterns of (a) undoped BFO, (b) bulk BDFO, (c) BDFO1, (d) BDF02 and (e) BDF03. Here the observed data are represented by red circles ($Y_{obs}$), the black line represents the calculated pattern ($Y_{cal}$) and the blue bottom curve is their difference ($Y_{obs}$-$Y_{cal}$), the colored bars correspond to Bragg positions of different phases of the corresponding samples.}\label{Fig:1}
\end{figure}

The Rietveld refined XRD patterns of the BFO bulk, BDFO bulk and its nano counterparts respectively are shown in  Fig.~\ref{Fig:1}. The BDFO nanoparticles sonicated for 30, 60 and 90 minutes are denoted by BDFO1, BDFO2 and BDFO3 respectively. FULLPROF package \cite{fullprof} was used to perform Rietveld refinement for quantification of crystallographic phases of these materials. The structural parameters obtained with the help of the refinement along with crystallographic phases (in wt\%) of BFO, BDFO bulk and nano samples are listed in supplemental Table 1 \cite{supp}. The major phase in all the samples was found to be of rhombohedral \textit{R3c} type crystal structure. Along with the major phase ($78.42\%$), the impurity phases Bi$_2$Fe$_4$O$_9$ ($9.47\%$) and Bi$_{25}$FeO$_{40}$ ($7.05\%$) were present in BFO. However, most of these secondary phases were removed with the substitution of Bi by Dy. The BDFO bulk comprised mostly of the characteristic rhombohedral \textit{R3c} phase ($93.98\%$), with reduced impurities. The phase purity of the ultrasonically prepared nanoparticles was almost unchanged for a sonication time of up to 60 mins as evidenced from Rietveld refinement (supplemental Table 1 \cite{supp}). Again, the enlarged view of the XRD patterns around the (104) and (110) peaks are depicted in supplementary Fig.~S1 \cite{supp} from which it can be clearly observed that the peaks shift to higher angles due to Dy substitution. In fact, this angle shifting is an indication of the shrinkage of unit cells of the Dy doped samples as a result of lower ionic radius of Dy$^{3+}$ compared to Bi$^{3+}$ \cite{ref13}. From Rietveld refined XRD patterns, the calculated lattice parameters a and c of the major phases were listed in Table \ref{Table:xrd1}. The XRD results confirmed that the substitution of Dy into the Bi site in BFO lead to changes in lattice parameters and crystal structure. Due to the substitution of Dy in Bi site, lattice parameters a and c (Table \ref{Table:xrd1}) were decreased compared to that of undoped BFO. The volumes of the major phases of the synthesized samples are also given in Table \ref{Table:xrd1}. 

 \begin{table}[t]
 	\centering
 	\caption{Calculated structural parameters from XRD}\label{Table:xrd1}
 	\begin{tabular}{|l|l|l|l|l|l|l|}
 		\hline
 		Sample & a=b & c & Volume & FWHM & Particle & strain\\ 
 	 & $\AA$ & $\AA$ & $\AA$$^3$ &(deg.) &  size (nm) & \\   \hline
 	 	BFO&5.5775(1)&13.8663(1)& 373.57&0.0104&&0.0241\\ \hline
 		BDFO&5.5635(8)&13.8191(74)& 370.44&0.0237&&0.0241\\ \hline	
 		BDFO1&5.5586(7)&13.8103(18)& 369.56&0.0256&35&0.0260\\ \hline
 		BDFO2&5.5616(12)&13.8136(23)& 370.01&0.0259&28&0.0382\\ \hline
 		BDFO3&5.5613(17)&13.8149(73)& 370.04&0.0260&24&0.0392\\ \hline  
 	\end{tabular}
 \end{table} 
  
  However, for the direct sonication probe dispersion technique \cite{probe}, a tendency of conversion from crystalline to amorphous phase was observed, indicated by the gradual decrease of the intensity of (104) and (110) peaks with increasing sonication time, demonstrated quite vividly in Fig.~\ref{Fig:1} (c--e). Beside the reduction of peak intensity in XRD patterns (Fig.~\ref{Fig:1} (c-e)), another indication of amorphization of the nanoparticles fabricated by ultrasonication is peak broadening. To investigate this effect, the full width at half maxima (FWHM) values of the peak at (012) are presented in Table \ref{Table:xrd1}. While performing this calculation, instrumental peak broadening factor was taken into consideration. For this purpose, the XRD peaks of amorphous SiO$_{2}$ were analyzed. 
 It was expected to have a flat pattern of SiO$_{2}$ with no distinct peaks since it is an amorphous material. However, due to instrumental peak broadening, some peaks were observed. So, it was important to perform necessary corrections \cite{inbroad} in the peak broadening of the synthesized samples with the aid of the reference profile. 
 
 \begin{figure}[hh]	
 		\centering
 	 	\includegraphics[width=3.2in]{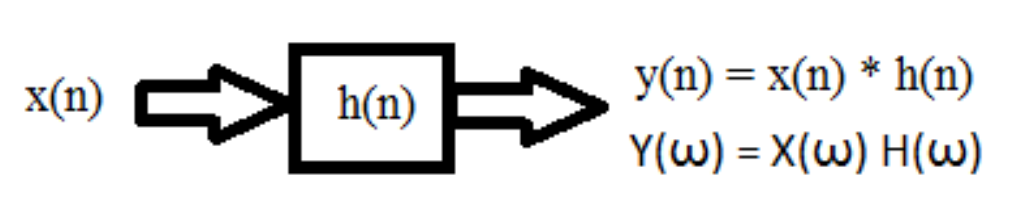}\\
 	 	\caption{Block diagram of a system with transfer function h(n).}\label{Fig:blckdia}
 	 \end{figure}
Since XRD machine, like any other system, has a certain transfer function, it introduces its own distortion effects on the actual XRD patterns of the samples. As a result, our obtained response is in fact the convoluted result of the desired intrinsic sample profile along with the instrumental broadening profile \cite{peakbroad}. Thus, despite simplified subtraction methods to negate this unwanted effect have been adopted assuming the shape profiles to be Cauchy--Cauchy (CC), Gaussian--Gaussian (GG) or Cauchy--Gaussian (CG) \cite{inbroad}, deconvolution operation is more suitable for accurate estimations. Let, the transfer function of the system is h(n), our expected XRD profile is x(n) and the obtained profile from the sample is y(n). Then, we may write-
 \begin{equation}
 y(n) = x(n) * h(n)
 \end{equation}
 where `*' represents convolution operation. This representation is shown in Fig.~\ref{Fig:blckdia}. The estimation of h(n) is performed from the response using the amorphous material as reference. Then the Fourier transforms are taken and we get \cite{peakbroad}-
 \begin{equation}
 X(\omega) = Y(\omega) / H(\omega)
 \end{equation}
 where, $X(\omega)$, $Y(\omega)$ and $H(\omega)$ are the Fourier transforms of x(n), y(n) and h(n) respectively. After that, performing inverse Fourier transform, we get an estimation of the actual sample response, x(n). Thus, correction for instrumental peak broadening is carried out and corrected parameters are calculated subsequently.
 The results in Table \ref{Table:xrd1} demonstrate the corrected broadening of the (012) peaks of BFO, BDFO, BDFO1, BDFO2 and BDFO3. The broad and diffuse peaks of Fig.~\ref{Fig:1} (c-e) once again indicate the tendency of crystalline to amorphous like phase transition \cite{ref210} of ultrasonically prepared nanoparticles. From here, we may conclude that crystallinity of the samples gets affected while formation of nanoparticles by using ultrasonic energy. However, the improvement in terms of phase purity from BDFO bulk to nano is not remarkable. Thus, the bulk samples provide satisfactory impurity phase reduction while preserving the crystalline structural attributes. 
 
 To further explore the structural properties of the samples, we calculated the particle size using modified Scherrer equation \cite{ref15} from the XRD peaks as given in Table \ref{Table:xrd1}. 
 As expected, the size of the ultrasonically prepared particles decreases with sonication time. Furthermore, we calculated the microstrain of the synthesized samples. From the values of Table \ref{Table:xrd1}, we observe that with increasing sonication period, the strain gradually increases. This once again reinforces the distortion in crystallinity induced by energy impact from ultrasonication for a longer duration. Again, we note that strain increases with decreasing particle size in the BDFO samples. This implies that with particle size reduction, the amount of lattice mismatch and dislocations increased. Such results are consistent with findings reported in \cite{ref1501}--\cite{ref1502}.
 
 To confirm the presence of expected amount of mass percentage of the atoms in synthesized Bi$_{0.9}$Dy$_{0.1}$FeO$_3$ samples, we performed Energy Dispersive X-ray spectroscopy (EDS). Supplementary Fig.~S2 \cite{supp} shows EDS spectrum of the bulk Bi$_{0.9}$Dy$_{0.1}$FeO$_3$ sample and from the data of table in the figure, it is verified that the synthesized sample contains the expected amounts of Bi, Fe, Dy and O.

 \subsection{Morphological Study}
\begin{figure}[h]	
  	\centering
  	\includegraphics[width=3.3in]{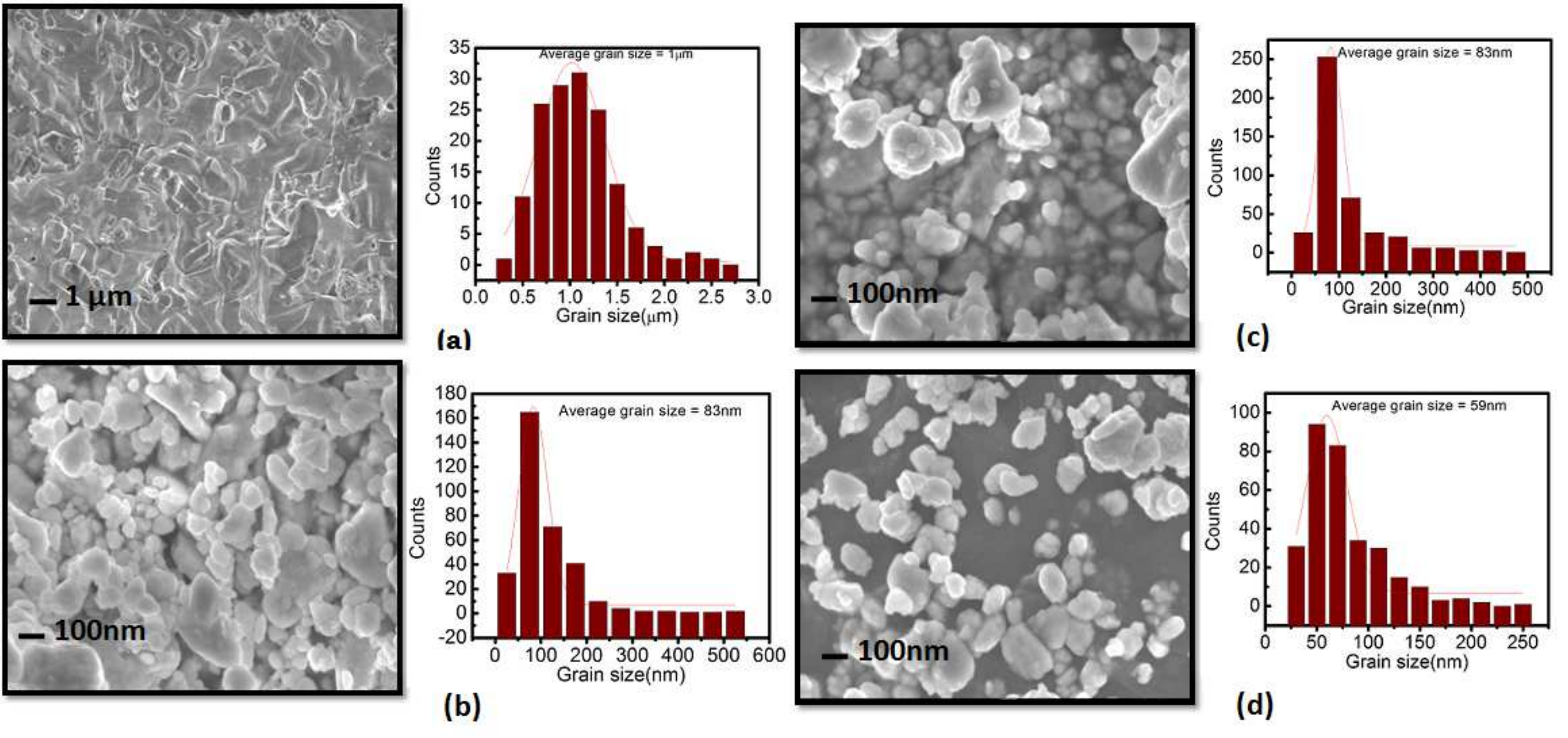}\\
  	\caption{FESEM images and their  corresponding histograms of (a)BDFO, (b) BDFO1, (c) BDFO2 and (d) BDFO3 samples. }\label{Fig:3}
  \end{figure}

 To investigate the surface morphology and  microstructure of the synthesized materials, FESEM imaging was carried out. Figure ~\ref{Fig:3} shown FESEM images and their corresponding histograms of the BDFO bulk, BDFO1, BDFO2 and BDFO3, respectively. Previous investigation showed the grain size of the bulk BFO to be 5$\mu$m$\sim$15$\mu$m \cite{ref10}. However, from the FESEM image, Fig.~\ref{Fig:3} (a), average grain size of BDFO bulk ceramic material was 1$\mu$m. The reduction of the grain size of the bulk material may be attributed to the reduction of oxygen vacancies with the addition of Dy ions into BFO \cite{ref15_1}. It  has  been  suggested  that  the  grain  growth depends  upon  the  concentration  of  oxygen  vacancies and diffusion  rate  of  the ions  \cite{ref20}.  Due  to  highly  volatile  nature  of  Bi,  its evaporation  causes  large  number  of  oxygen  vacancies  in pure BFO.  This  makes  it  easier for  the  ions  to  diffuse,  resulting in  a  very  large  grain  size  as  compared  to  the Dy doped BFO sample. But, when the Dy ion is substituted within BFO at Bi site, some amount of volatile Bi gets substituted. Again, as mentioned previously, the bond enthalpy of Dy-O (607 $\pm$ 17 kJ/mol) is stronger than that of Bi-O (337 $\pm$ 12.6 kJ/mol) \cite{ref10b} which implies that a higher energy is required to cause breaking of Dy-O bond resulting in a lower probability of oxygen vacancies. As a result, Dy doping constrains volatilization of Bi and the concentration of  oxygen vacancies is lowered \cite{ref15_1}. 
 
 The ultrasonically prepared particles calculated from the histograms for BDFO1, BDFO2 and BDFO3 materials are 75-150, 75-100 and 50-100 nm, respectively. It is observed that the average particle size decreases with the increase of sonication time as shown in Fig.~\ref{Fig:3} (b-d) for 30, 60 and 90 mins sonication, respectively. This is due to the fact that ultrasound irradiation generates many localized hot spots with the particles within the solution \cite{ref18} and during the process, the implosive collapse of the bubble causes an inward rush of liquid known as `microstreaming' in which high velocity is produced. Such cavitational collapse during sonication in solids leads to microjet and shock-wave impacts on the surface \cite{ref18_1}. These phenomena together with interparticle collisions might have played a role in particle-size reduction \cite{ref18_2}. The calculated size of the ultrasonically prepared particles using Scherrer's formula as shown in Table \ref{Table:xrd1} is 
 smaller than the value observed from SEM images. The large particle size determined by electron microscopy images compared than that calculated by Scherrer equation has also been reported in previous investigations \cite{ref901}---\cite{ref902}
 and this is an indication of the agglomeration of the particles. Again, it has been reported that ultrasonic cavitation can generate extremely high local temperatures
 and pressures along with rapid heating and cooling rates \cite{ultrasonication}. Thus the agglomerations can be broken down utilizing the effects of ultrasound \cite{ultrasonication}, resulting in reduction of particle size. 

In our previous investigation \cite{ref11}, Gd and Ti co-doped BiFeO$_3$ nanoparticles were produced using ultrasonic energy through bath sonication (indirect process) from their bulk powder materials. Noticeably, in our current investigation for ultrasonic probe dispersion (direct process) sonication, 30 minutes of sonication was enough to produce nanoparticles within the range of 75 nm to 150 nm, whereas, within the same timeframe,  bath  sonication (indirect) process was able to produce 150 to 250 nm grains only. Although the doping elements in this investigation and those of Ref.~\cite{ref11} are different, we get an idea of the scale of particle size reduction for the two ultrasonication procedures from these two cases. Since for the direct process, lesser time is required to produce the same amount of reduction in particle size, we may infer that for ultrasonic probe dispersion (direct process), higher mechanical energy is bombarded at the particles from the ultrasonic wave compared to bath technique. Hence, probe dispersion technique was used in this investigation.

\subsection{Magnetic Characterization}
Having studied the structural properties of the synthesized materials, this subsection is dedicated to the investigation of the magnetic properties of BDFO bulk material and their nano counterparts. 
\subsubsection{Field dependent magnetization}
\begin{figure}[hh]	
	\centering
	\includegraphics[width=3.5in]{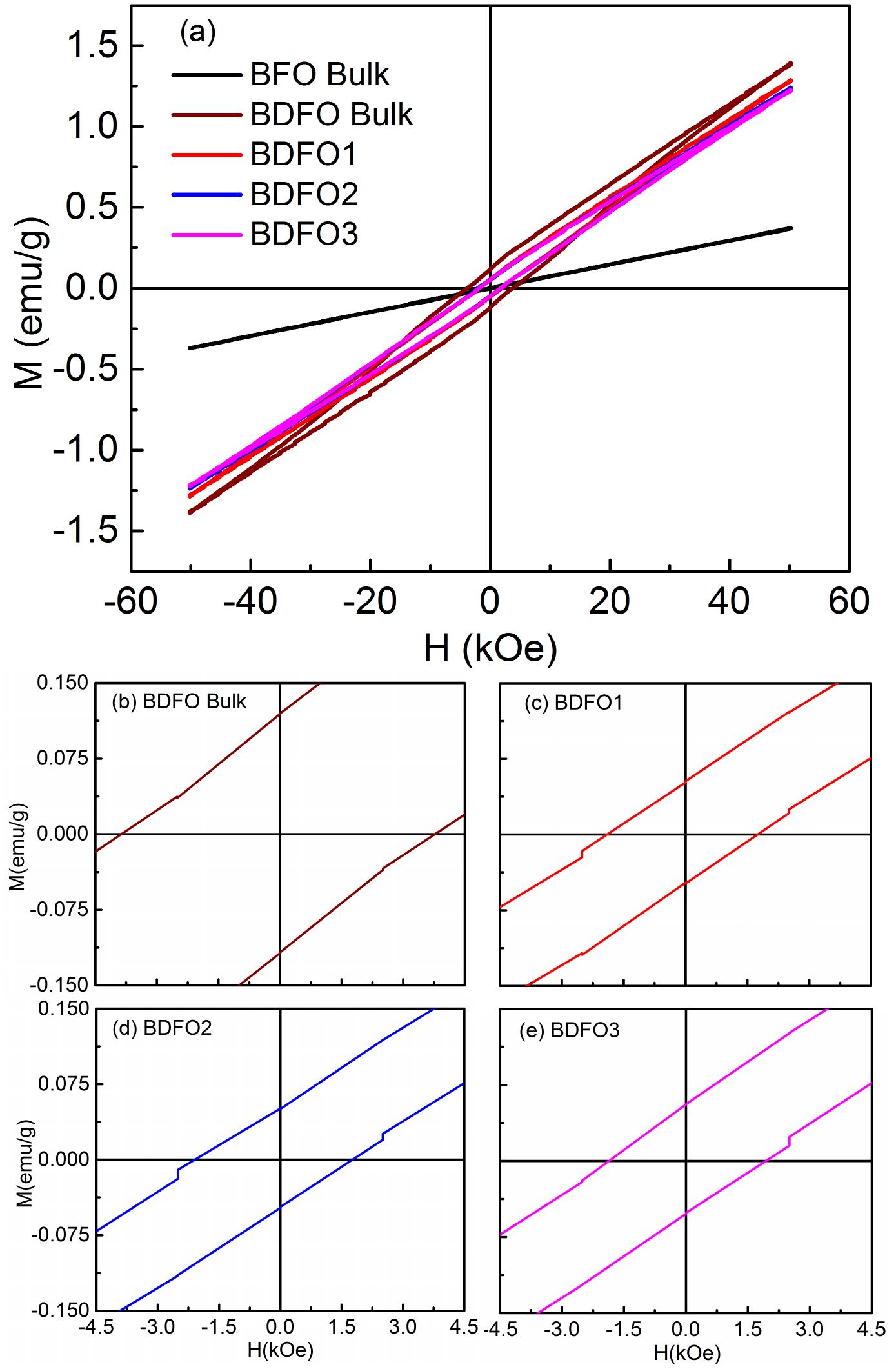}\\
	\caption{(a) Room temperature M-H hysteresis loops of undoped bulk BFO, BDFO bulk sample and
		ultrasonically prepared nanoparticles. (b-e) The enlarged view of bulk BDFO, BDFO1, BDFO2 and BDFO3, respectively.}\label{Fig:7}
\end{figure}

In order to obtain the magnetic characteristics of the samples, field dependent magnetization measurements were carried out. The M-H hysteresis loops of BDFO bulk
sample and its ultrasonically prepared nanoparticles measured at room temperature with
an applied magnetic field of up to $\pm$50 kOe are shown in Fig.~\ref{Fig:7} (a). For comparison M-H loop of undoped bulk BFO ceramic is also carried out and inserted in Fig.~\ref{Fig:7} (a). 
The enlarged view of this hysteresis loops are shown in Fig.~\ref{Fig:7} (b-e) for BDFO bulk, BDFO1, BDFO2 and BDFO3, respectively. Unlike undoped bulk BFO as shown in 
Fig.~\ref{Fig:7} (a), the BDFO bulk sample exhibited weak ferromagnetic nature with many folds
higher values of the magnetization compared to that of undoped BFO \cite{ref11}. 

\begin{table}[h]
	\centering
	\caption{Calculated values of M$_r$, H$_c$ and
		H$_{EB}$ for the synthesized ceramics.}\label{Tab2}
	\begin{tabular}{|l|l|l|}
		\hline
		Sample & Coercive Field, &  Remanent \\
		Specification & H$_c$ (Oe) & Magnetization, \\
		 & (Oe) & M$_r$ (emu/g)	  			  		\\ \hline
		BFO bulk&132&0.001\\ \hline
		BDFO bulk&3793& 0.118\\ \hline	
		BDFO1&1761 &0.053\\ \hline
		BDFO2&1822  &0.050\\ \hline
		BDFO3&1835& 0.050\\ \hline
	\end{tabular}
\end{table} 

From these hysteresis loops of Fig.~\ref{Fig:7} (a), formulas used to obtain quantitative measures of the magnetic parameters like (i) coercive fields (H$_c$) and (ii) remanent magnetization (M$_r$) were: $H_c = (H_{c1}-H_{c2})/2$, where H$_{c1}$ and H$_{c2}$ are the left and right coercive fields \cite{ref10, ref25} and M$_{r}$ = $|$(M$_{r1}$-M$_{r2}$)$|$/2, where M$_{r1}$ and M$_{r2}$ are the magnetization with positive and negative points of intersection with H = 0, respectively \cite{ref7}. 
Calculated values of M$_{r}$ and H$_c$ for the samples are inserted in Table \ref{Tab2}. 

The observed value of M$_r$ of BDFO bulk sample was
found to increase significantly compared to that of BFO. This M$_r$ is 
higher than the reported values for some A-site doped and co-doped BFO bulk ceramics \cite{ref26},\cite{ref27}. It is generally expected that due to Dy doping, the spin cycloid structure of
BFO would be destroyed. As a result, the latent magnetization locked within the
cycloid structure is released \cite{ref28}. Previous investigations have reported further enhancement of magnetization
values \cite{ref26},\cite{ref27} when nanoparticles were synthesized and characterized from their
corresponding bulk compositions of similar materials. This induced us to perform
further characterization of the ultrasonically prepared nanoparticles of BDFO with anticipation of
improved magnetic properties. However, the M$_{r}$ of these
ultrasonically prepared BDFO nanoparticles was, in fact, found to decrease
compared to that of BDFO bulk material. It should also be noted that sonication for a longer time actually does not change M$_{r}$ significantly as can be seen from Table \ref{Tab2}. The high crystal quality of the bulk BDFO
sample as evidenced by low microstrain and low leakage
current (described later on in subsection D) is probably the
reason behind the higher magnetization of this bulk ceramic
material. We also think that the amorphization of
the nanoparticles prepared by the probe generated ultrasonic
energy as evidenced from XRD analysis causes the decrement of M$_r$ in
corresponding nanoparticles. 

It is worth mentioning that the enhanced $M_r$ value for BDFO is quite significant even when compared to Gd-doped BFO with $M_r$ = 0.065 emu/g \cite{ref29}. So, the BDFO bulk sample in this investigation provides superior magnetization performance compared to their corresponding nanoparticles as well as some other similar doped ceramics. To further investigate the improved magnetic properties of BDFO bulk over its nano counterparts, we compare the obtained results with results reported in \cite{ref17}, where Dy doped BiFeO$_3$ nanoparticles were synthesized by a low temperature co-precipitation method. For the 10\% Dy doped BiFeO$_3$ nanoparticles, the reported $M_r$ value  was $\sim$ 0.02 emu/g in Ref. \cite{ref17}, whereas, our obtained $M_r$ for BDFO bulk material is 0.118 emu/g. Again, for the same composition synthesized by chemical co-precipitation \cite{ref17}, the obtained maximum magnetization was 0.47 emu/g at 15 kOe. At the same level of applied field, we obtained maximum magnetization of 0.58 emu/g for BDFO bulk material. So, clearly the enhanced magnetic properties of BDFO bulk ceramic are consistent with nanoparticles produced by other synthesis procedure as well. Moreover, these values are higher than that of undoped BFO (0.11 emu/g) \cite{ref29a} as well as few other published results on Dy doped BFO ceramic \cite{ref26}.  

From Table \ref{Tab2}, it is observed that there is a significant increase in coercivity in the bulk sample from BFO to BDFO bulk. Such enhancement may be due to the increase in anisotropy of the samples. 
Similar results have been observed by Rai \textit{et al.}, where
occurrence of high H$_c$ values were reported with La doping \cite{ref1504}. Large coercivity
of the order of 1–-2 T at room temperature have also been reported for
other dopant systems \cite{ref1505}. Among the BDFO nanoparticles, we observed an increasing trend in H$_c$ values with decreasing particle size. Such variation may be explained based on domain structure and surface and interface anisotropy of the crystal. In case of nanoparticles, coercivity depends on two factors mostly--(a) spin rotation through which the magnetization will be reversed rather than through the motion of domain walls and (b) shape anisotropy \cite{ref1506}. In order to minimize the large magnetization energy, spontaneous breaking up of a crystallite into multiple of domains will occur considering it was a single domain to start with. The ratio of the energy before and after division into domains varies as $\sqrt{D}$ \cite{ref1507}, where $D$ is the particle size. So, with decrease in D, the tendency of energy reduction due to splitting into domains also decreases suggesting that the crystallite prefers to remain single domain for quite small D. As a result, particle size reduction lowers the probability of domain splitting and for a single domain like crystallite, higher amount of field will be required for magnetization reversal. This in turn, explains the increase in H$_c$ values with decreasing particle size for BDFO nanoparticles.
\begin{figure}[hh]	
	\centering
	\includegraphics[width=3.2in]{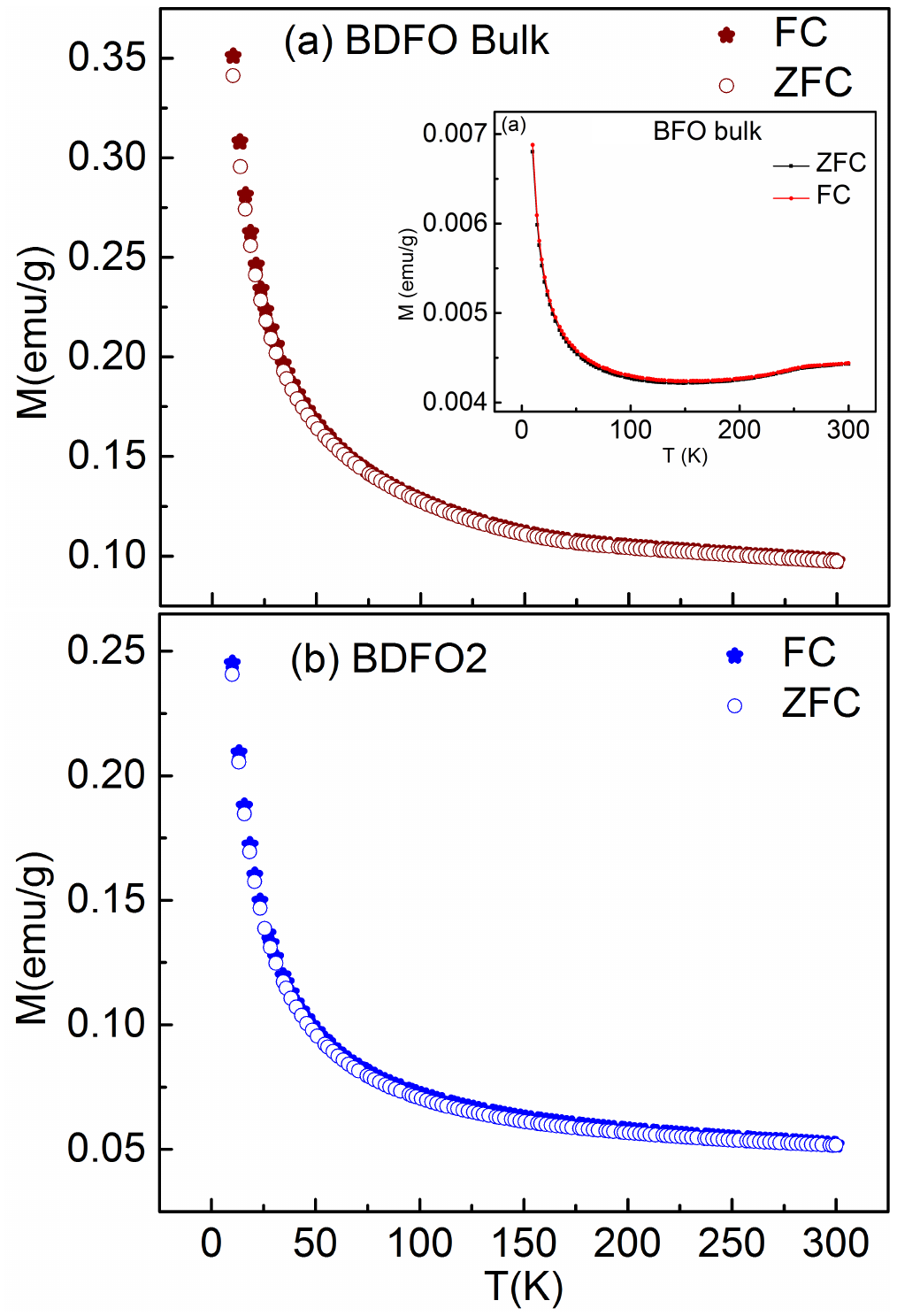}\\
	\caption{Temperature dependent ZFC and FC magnetization (M-T
		curves) of (a) bulk BDFO ceramic and (b) BDFO2 nanoparticles. Inset contains the same for undoped BFO bulk.}\label{Fig:9}
\end{figure}

In the subsequent investigation, to compare the results with nano
sized samples, we chose to experiment with BDFO2 (60 min
sonicated sample) because for BDFO1, the size of some of the particles was
large (Fig. ~\ref{Fig:3} (b)) and for BDFO3, there was higher amorphization.
Thus to obtain the best possible characteristics of the nanoparticles
among the synthesized samples for comparison purpose with
BDFO bulk, BDFO2 was picked.

\subsubsection{Temperature dependent magnetization}
At this stage of the investigation, temperature dependent zero field cooled (ZFC) and field cooled (FC) magnetization measurements under a magnetic field of 500 Oe were carried out to further explore the magnetic properties of BDFO bulk material and corresponding nanoparticles. The ZFC process comprised of initial cooling of the sample from 300K to 5K and collection of data while heating in the presence of the
applied field. On the contrary, while experimenting in the FC mode,  data were collected during cooling in
the presence of 500 Oe magnetic field \cite{ref29}.

In Figs. \ref{Fig:9} (a) and (b), we have demonstrated the magnetization vs. temperature (M--T) curves of the
BDFO bulk and BDFO2 nanoparticles, respectively. As can be seen, both ZFC and FC magnetization curves
overlap with each other throughout the temperature range under investigation
indicating that no magnetic transition has occurred. In both cases, the
magnetization value gradually increased with decreasing temperature. The
magnetization increases slowly as temperature is decreased up to 150 K and then
sharply rises up with further decreasing temperature. The steep increase of
magnetization particularly below 50 K in Figs. \ref{Fig:9} (a) and (b) indicates the weak
ferromagnetic nature of this material system at sufficiently low temperature. Similar
result was observed by Lu \textit{et al.} \cite{ref30} which suggests that
the sample has retained its intrinsic property where no phase or glass transition has
taken place. This, in fact, is perhaps most likely due to enhanced phase purity in the
synthesized samples. Also in the inset of Fig.~\ref{Fig:9} (a), the M-T curves of undoped BFO are plotted. For this sample,
Basith \textit{et al.} reported an anomaly around 264 K \cite{ref29} where a neck in the curve was
found. However, the bulk BDFO sample and nano sized BDFO2 under scrutiny did not show any such anomaly. The transition that is usually observed in BiFeO$_3$ around 264 K is due to the impurity phase (Bi$_2$Fe$_4$O$_9$). This impurity phase has antiferromagnetic transition around 260 K \cite{ref1508}. So absence of this peak in the samples implies that suppression of this Bi deficient phase was achieved through Dy substitution.

\subsection{Electrical Measurements}
 \begin{figure}[hh]	
 	\centering
 	\includegraphics[width=3.2in]{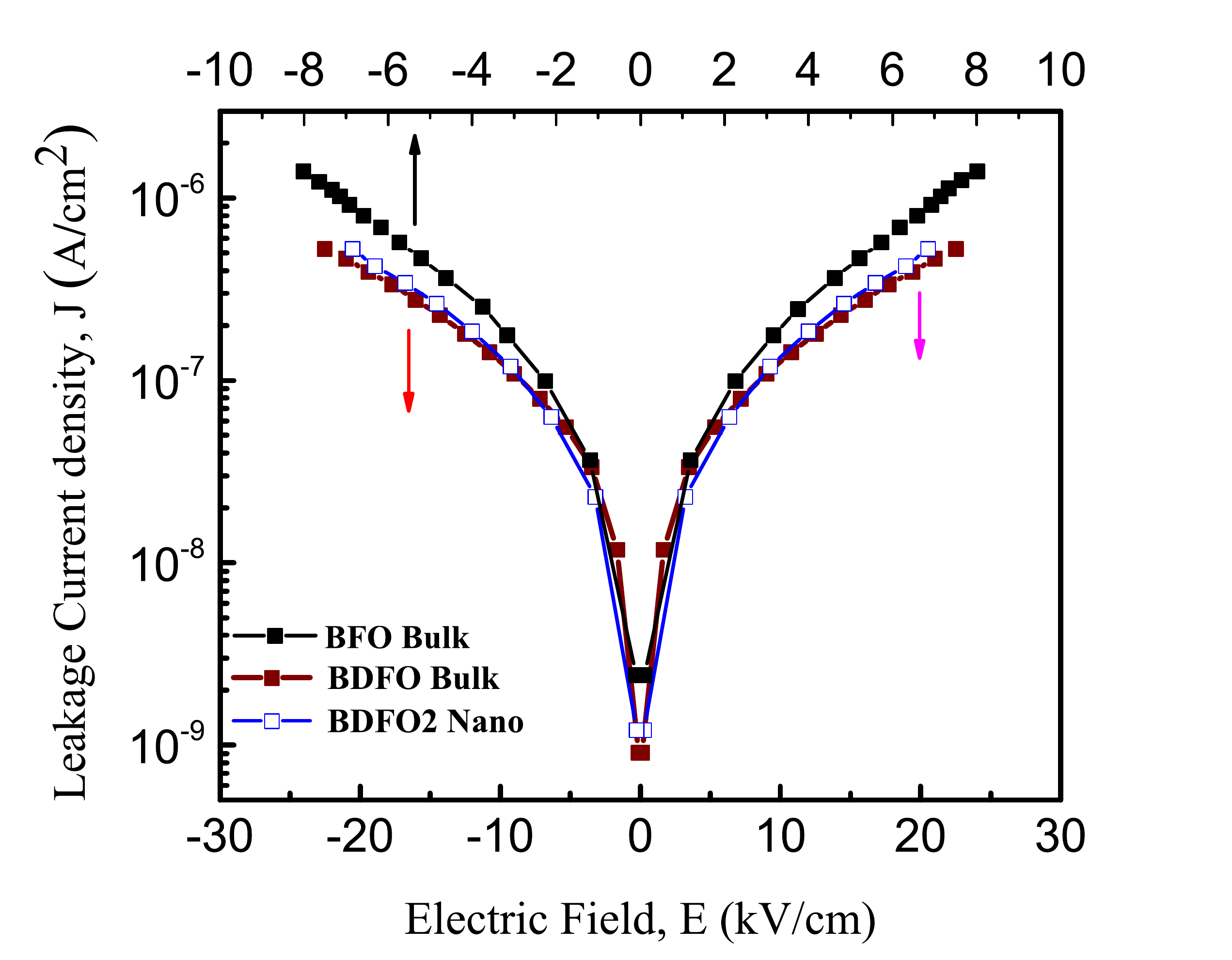}\\
 	\caption{Leakage current density of BFO, BDFO bulk materials and their nanoparticles sonicated for 60 minutes as a function of applied electric field. The upper x-axis values correspond to bulk BFO.}\label{Fig:4}
 \end{figure}
To compare the electrical properties of BDFO bulk along with corresponding nanoparticles and undoped bulk BFO material, leakage current density, J versus electric field, E measurements were performed for an applied field of up to $\pm20$ kV/cm as shown in Fig.~\ref{Fig:4}. Notably, for the ferroelectric measurements, BDFO2 was chosen as a representative of the ultrasonically prepared nanoparticle to be explored and compared with the bulk sample for reasons explained in the previous subsection. It is clear that leakage current density reduced significantly from BFO to BDFO bulk. This phenomenon may be explicated from the fact that very few impurity phases were present in BDFO bulk sample as was evidenced from Rietveld refinement. However, when we experimented with BDFO2, the leakage current density was slightly higher over the whole range of applied field compared to BDFO. The lower leakage current density in the BDFO bulk compared to undoped BFO ceramic 
\begin{figure}[hh]	
	\centering
	\includegraphics[width=3.2in]{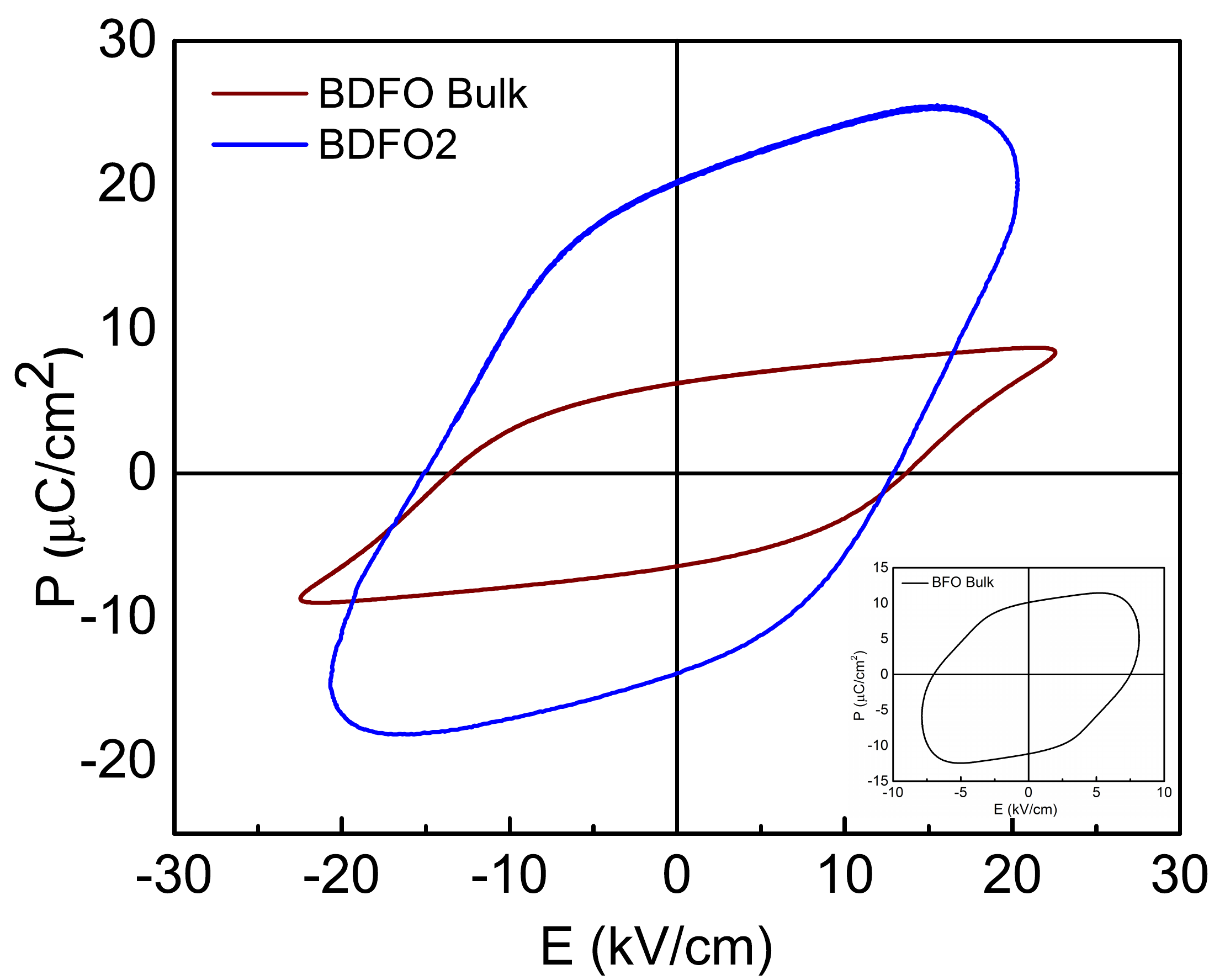}\\
	\caption{Room temperature polarization vs. electric field loops of BDFO bulk materials and their nanoparticles prepared by ultrasonication sonicated for 60 minutes. The inset shows the P--E loop of BFO bulk for comparison.}\label{Fig:5}
\end{figure} 
might be due to the suppression of oxygen vacancy related defects by Dy doping in BFO ceramic. Again, the improved leakage current density performance of the BDFO bulk over undoped BFO and corresponding nanoparticles provides a very notable result in this case. This could be very promising for applications with low leakage current requirements where we no longer need to go for nanoparticles which involve multiple preparation steps. Rather the bulk BDFO sample would provide us acceptable solution with improved properties. Although the difference between leakage current densities of BDFO bulk and their nanoparticles is small, however they are much lower than the parent BFO ceramic. As a result, the  polarization versus electric field (P--E) loops of BDFO bulk ceramic and their corresponding nanoparticles are expected to exhibit enhanced ferroelectric properties. To confirm this, next we investigate the P--E hysteresis loop measurements.

Figure \ref{Fig:5} shows the P--E loops for BDFO and BDFO2 carried out at room temperature with a frequency of 50 Hz. The P--E loop of bulk BFO is also given in the inset for comparison. We found that the remanent polarization (P$_{\text{r}}$) and the coercive field (E$_{\text{c}}$) of BDFO2 are higher compared to that of bulk BDFO. 
The remanent polarization P$_{\text{r}}$ for BDFO was 6.24 $\mu$C/cm$^2$ while for BDFO2 it increased to 16.98$\mu$C/cm$^2$.

The larger polarization in BDFO2 nanoparticles may be associated with higher leakage current, reduced electrical resistivity and hybridization of the 6s$^{2}$ lone pair electrons of Bi$^{3+}$ with the 2s/2p orbitals of O$_{2}$ \cite{ref17}. As we can see, the small difference in leakage current between BDFO bulk and BDFO2 nanoparticles created a large impact in P--E hysteresis. In bulk BDFO ceramic, P$_{\text{r}}$ is lower, however, the loop is nearly saturated. The P-E
hysteresis loops for BFO multiferroics are often dominated by large 
leakage current affected by mixed valence of 
Fe$^{2+}$ and Fe$^{3+}$ ions, or by oxygen vacancy related defects, or by both \cite{ref21}. The interpretation of ferroelectric behavior of materials with such P-E loops can often be misleading which may lead to erroneous conclusions. It has been asserted in Ref. \cite{ref22} that true
ferroelectrics exhibit saturation in
polarization and have concave region in P-E loop. We observed that BDFO bulk provides saturated polarization while BFO bulk and BDFO2 nanoparticles do not, indicating more reliable ferroelectric characteristics for BDFO bulk. The oxygen vacancies due to volatility of Bi and hopping of itinerant electrons between Fe$^{2+}$ and Fe$^{3+}$ ions result in
high conductivity and leakage current in undoped BFO. Hence we may anticipate that
replacement of highly volatile Bi with Dy reduces the concentration
of oxygen vacancies giving reduced leakage current and thus improves ferroelectric properties. Notably, the ferroelectric properties of Dy doped bulk BFO ceramic were improved compared to that of some other rare earth doped e.g. La or Gd doped BFO bulk ceramics \cite{ref601}---\cite{ref602}.


\subsection{Optical Characterization}
For determining the optical band gap of the synthesized BDFO bulk
materials and ultrasonically prepared nanoparticles (BDFO2), UV visible diffuse reflectance spectra (DRS) were recorded at room temperature. Initially,
40 mg of BDFO bulk materials and nanopowders were pressed with 2 g of
Barium Sulphate (BaSO$_4$) salt into a thin translucent disc, as the BaSO$_4$ salt has no
significant light reflection in visible light region and also transparent to near
ultraviolet to the long-wave infrared wavelengths region \cite{ref31}. 

\begin{figure}[t]	
	\centering
	\includegraphics[width=3.2in]{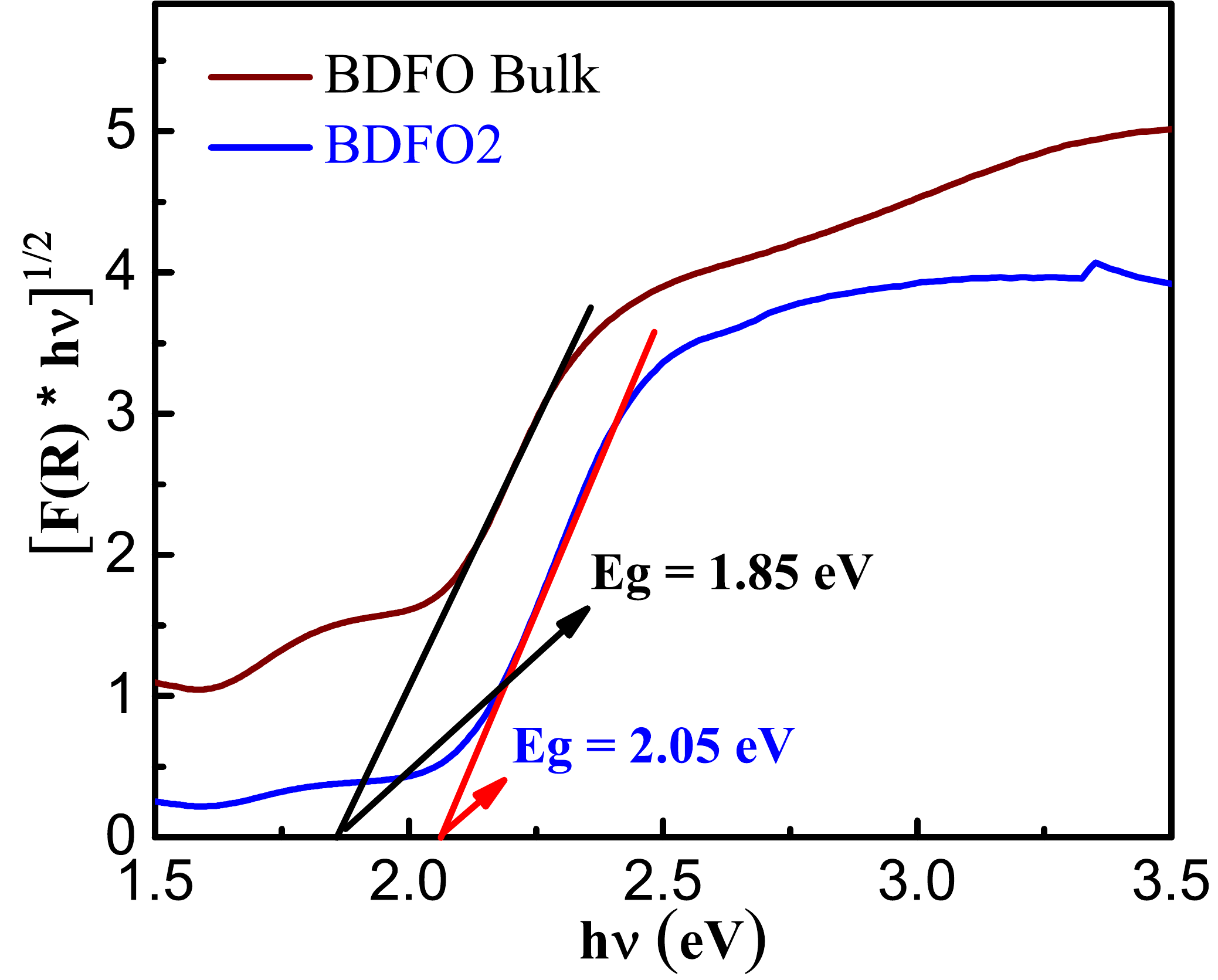}\\
	\caption{Bandgap energy estimation from diffuse reflectance spectra of BDFO and BDFO2.}\label{Fig:12}
\end{figure}

The experimental results of the room temperature UV-visible diffuse reflectance spectra for
BDFO bulk material and BDFO2 nanoparticles are depicted in Fig.~S3 of supplementary information \cite{supp}. From the results, we see that for the whole range of wavelengths, BDFO provides lower reflectance than the nanoparticles. This is an indication of BDFO bulk being the more suitable material compared to BDFO2 nanoparticles in photocatalytic applications where we need a larger absorption of light such as in solar hydrogen production.

Next, we determine the bandgap of the synthesized materials from the diffuse reflectance spectrum using Kubelka-Munk
function which is proportional to the extinction coefficient \cite{ref4c},
\begin{equation}
F(R)=\frac{(1-R)^2} {2R}.
\end{equation}
The bandgap energy for the photocatalysts can be calculated by using the following equation \cite{ref4c},
\begin{equation}
F(R)^{*}h\nu=A(h\nu-E_{g})^n,
\end{equation}
where, h$\nu$, A and E$_g$ denote the energy, proportionality constant and bandgap energy respectively.

Fig.~\ref{Fig:12} displays $[F(R)*h\nu]^{1/2}$ vs $h\nu$ curves to calculate the band gap of BDFO and BDFO2. The band gap is calculated by extrapolating the linear portion of the
curve to the x-axis. The calculated optical band gap values are 1.85 eV and 2.05 eV
for bulk BDFO and BDFO2 nanoparticles, respectively. The obtained values of the band gap for
these samples are smaller compared to the reported values for pure and doped BFO bulk ceramics and BFO nanoparticles \cite{ref33}--\cite{ref35}, which is around 2.1 eV. Similar increase in bandgap with decrease in particle size was reported in \cite{ref65}. The absorption edge of the smaller BDFO2 particles underwent a blueshift which may be attributed to quantum-size effect \cite{ref66}. In general, in a single atom the bandgap is equal to the distance between ground state and first excited state, while in the bulk where we have a huge number of atoms, both levels are broadened. Such broadening leads to narrowing of the bandgap. In a nanoparticle, broadening is less than in the bulk due to smaller number of atoms present in the system, and a wider bandgap is obtained as result. Once again, we note that, the BDFO bulk provides a more suitable option for solar applications for capturing larger amount of light compared to the nanoparticles. Since, the bulk material possesses lower bandgap, the cut-off wavelength ($\lambda_{cut}$) of BDFO bulk is higher than that of BDFO2 nanoparticles as the bandgap energy is inversely proportional to the cut-off wavelength. The considerably lower bandgap of the bulk BDFO ceramic compared to that of undoped bulk BFO is notable since it is closer to the optimal bandgap (1.50 eV) for photovoltaic applications \cite{lowbandgap}. For solar applications, the energy of the incoming photon must be greater than the bandgap of the material in order to create electron-hole pairs. Thus, only photons up to $\lambda_{cut}$ can contribute in such cases. As a result of the higher $\lambda_{cut}$, bulk BDFO can absorb photons with a wider range of energy than the nano counterparts as well as undoped BFO.     
 
\section{Conclusion}
The structural, multiferroic and optical characterizations of BDFO bulk and corresponding nanoparticles were performed in this investigation. XRD results indicated that BDFO bulk and their nano counterparts possess similar phase purity. However, M-H measurements demonstrated higher remanent magnetization and coercivity for the bulk BDFO ceramic compared to those for corresponding nanoparticles. Leakage current density measurements also showed that  BDFO bulk sample had even lower leakage current compared to their corresponding nanoparticles. Interestingly, true saturation polarization was also observed only for Dy doped bulk BFO ceramic sample. Furthermore, bulk BDFO ceramic showed improved results compared to nanoparticles by displaying lower bandgap and larger absorption which are important for photocatalytic applications. Therefore, it can be inferred that Dy doped BiFeO$_3$ ceramic exhibited improved performances as a bulk material in comparison to its corresponding nanoparticles.  Notably, synthesis of multiferroic nanoparticles with improved properties involves multiple complex chemical, physical, and physiochemical interactions and it is indeed a great challenge to understand the synthesis mechanisms clearly. Thus, considering the experimental difficulties associated with synthesis of multiferroic nanoparticles, we may conclude that 10\% Dy doped BiFeO$_3$ material would be a promising bulk multiferroic ceramic with notable properties compared to its nano counterparts.

\section*{Acknowledgment}
This work was financially supported by Infrastructure Development Company Limited (IDCOL), Dhaka, Bangladesh. The Institute for Molecular Science (IMS), supported by Nanotechnology Platform Program (Molecule and Material Synthesis) of MEXT, Japan is acknowledged for providing SQUID facilities.


\end{document}